\documentclass{PoS}

\title{Differential cross sections for top-pair and single-top production}

\ShortTitle{Differential cross sections for top-pair and single-top production}

\author{\speaker{Nikolaos Kidonakis}%
         \thanks{This material is based upon work supported by the National Science Foundation under Grant No. PHY 1212472.}\\
        Kennesaw State University, USA\\
        E-mail: \email{nkidonak@kennesaw.edu}}

\abstract{I present higher-order results, based on NNLL resummation, for differential transverse momentum and rapidity distributions in processes involving top quarks. In particular results are presented for top-pair production and for single-top production in different channels.}

\FullConference{The European Physical Society Conference on High Energy Physics \\
                 18-24 July, 2013\\
                 Stockholm, Sweden}

\begin{document}

\section{Introduction}

QCD corrections at next-to-leading order (NLO) and higher orders in the strong coupling are known to be significant for top pair and single top production. An important set of these corrections are those 
from soft-gluon emission and they turn out to be numerically dominant. 
The analytical structure of the soft-gluon terms in the perturbative series involves logarithmic plus distributions
$[\ln^k(s_4/m_t^2)/s_4]_+$, with $m_t$ the top-quark mass, $k \le 2n-1$ for the $n$-th order corrections, and $s_4=s+t_1+u_1$ (with $s$, $t_1$, $u_1$ standard kinematical variables) the kinematical distance from partonic threshold, which is more general than absolute threshold (i.e. top quarks are not necessarily produced at rest and can have an arbitrarily large transverse momentum, $p_T$).

These soft corrections have now been resummed to  next-to-next-to-leading-logarithm (NNLL) level, and approximate NNLO results have been obtained for top-pair \cite{NKtop} and single top \cite{NKst} production. The calculations in \cite{NKtop,NKst} are all for double-differential cross sections, e.g. in $p_T$ and rapidity, using the moment-space resummation formalism in perturbative QCD.

\begin{figure}
\begin{center}
\includegraphics[width=10cm]{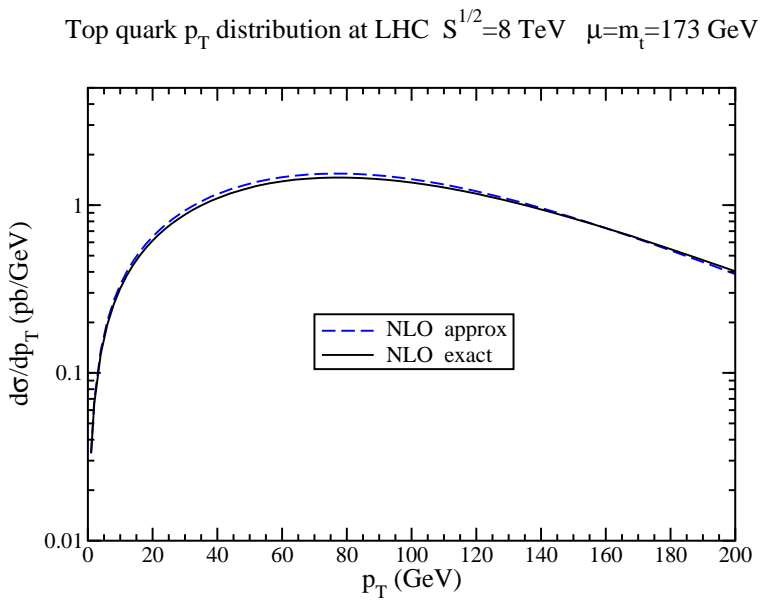}
\caption{Comparison of exact and approximate NLO results for the top-quark $p_T$ distribution in $t{\bar t}$ production at 8 TeV LHC energy.}
\label{ptcorr8lhcmplot}
\end{center}
\end{figure}

The threshold approximation works very well for LHC and Tevatron energies; in fact it is an excellent approximation for total and differential top quark cross sections: there is only $\sim$1\% difference between NLO approximate and exact cross sections, and similarly for differential distributions in $p_T$ and rapidity. The excellence of the approximation is expected to persist also at NNLO (as has been checked for total cross sections, see e.g. discussions in \cite{NKsnowmasseps13}).

In Fig. \ref{ptcorr8lhcmplot} we compare the approximate NLO result for the top quark $p_T$ distribution in $t{\bar t}$ production at 8 TeV LHC energy to the exact NLO \cite{NLOttbar} result. As can be seen in the plot the approximation is excellent and this shows that soft-gluon emission provides the dominant contribution. For the best prediction for differential distributions one must add the NNLO 
approximate corrections from NNLL resummation \cite{NKtop} to the exact NLO results \cite{NLOttbar}.

\section{Top quark distributions in $t{\bar t}$ production}

\begin{figure}
\begin{center}
\includegraphics[width=10cm]{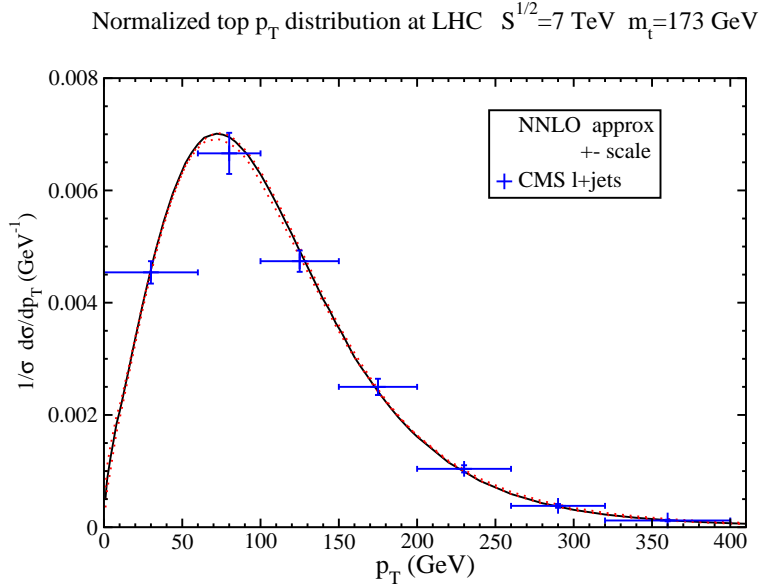} 
\caption{The approximate NNLO normalized top-quark $p_T$ distribution at 7 TeV LHC energy compared with CMS data \cite{CMS7} in the ${\ell}$+jets channel. The solid line is the central result with $\mu=m_t$ and the dotted lines show the scale variation.}
\label{pt7lhcnormCMSljetplot}
\end{center}
\end{figure}

We begin with top-antitop pair production. In our results we use the MSTW2008 NNLO parton densities \cite{MSTW}. In Fig. \ref{pt7lhcnormCMSljetplot} we show the normalized top quark $p_T$ distribution, $(1/\sigma) \, d\sigma/dp_T$, at approximate NNLO at 7 TeV LHC energy using a top quark mass of 173 GeV. The variation with renormalization and factorization scales over the interval $m_t/2 < \mu <2m_t$ is displayed, and the central result is with $\mu=m_t$. In addition, the CMS data \cite{CMS7} in the ${\ell}$+jets channel are also shown. The agreement with theory is excellent and is better than at NLO, as shown in \cite{CMS7}. The effects of the NNLO soft-gluon corrections is to soften the distribution; this is a natural extension of the fact that the NLO corrections (either soft-gluon alone or complete) soften the LO distribution. Similar agreement is found with CMS data \cite{CMS7} in the dilepton channel. Finally, more recent data from CMS at 8 TeV energy \cite{CMS8} show again a softening of the spectrum and excellent agreement with the NNLO approximate result.

\begin{figure}
\begin{center}
\includegraphics[width=9.5cm]{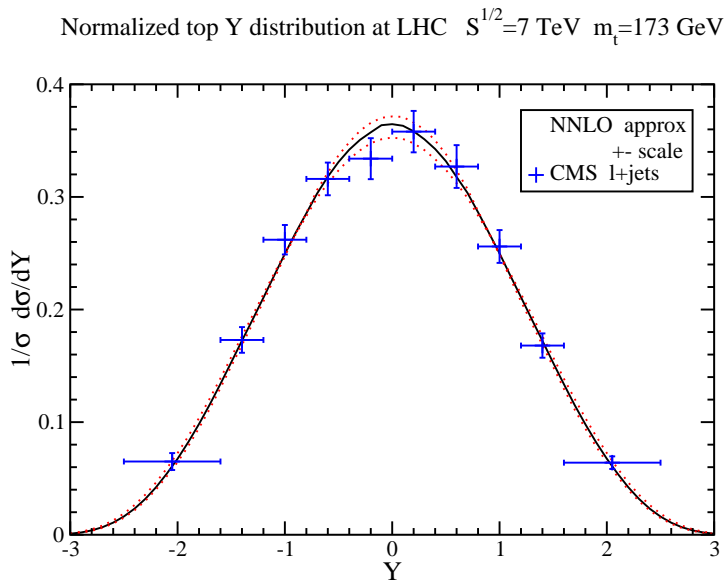}
\caption{The approximate NNLO normalized top-quark rapidity distribution at 7 TeV LHC energy compared with CMS data \cite{CMS7} in the ${\ell}$+jets channel. The solid line is the central result with $\mu=m_t$ and the dotted lines show the scale variation.}
\label{y7lhcnormCMSljetplot}
\end{center}
\end{figure}

In Fig. \ref{y7lhcnormCMSljetplot} we show the normalized top quark rapidity distribution, $(1/\sigma) \, d\sigma/dY$, at approximate NNLO at 7 TeV LHC energy using a top quark mass of 173 GeV. Again the variation with scale is displayed.
The agreement with CMS data at 7 TeV in the ${\ell}$+jets channel \cite{CMS7} shown in the plot is again excellent. Similar agreement is found with CMS data in the dilepton channel \cite{CMS7} and also at 8 TeV energy \cite{CMS8}.

\section{Top quark distributions in single-top production}

We continue with single-top production and present top-quark transverse-momentum distributions in $t$-channel and $tW$ production. The NLO differential distributions for single-top production have been known for over a decade \cite{NLOsingletop}. Here we show approximate NNLO results from NNLL resummation.

\begin{figure}
\begin{center}
\includegraphics[width=9cm]{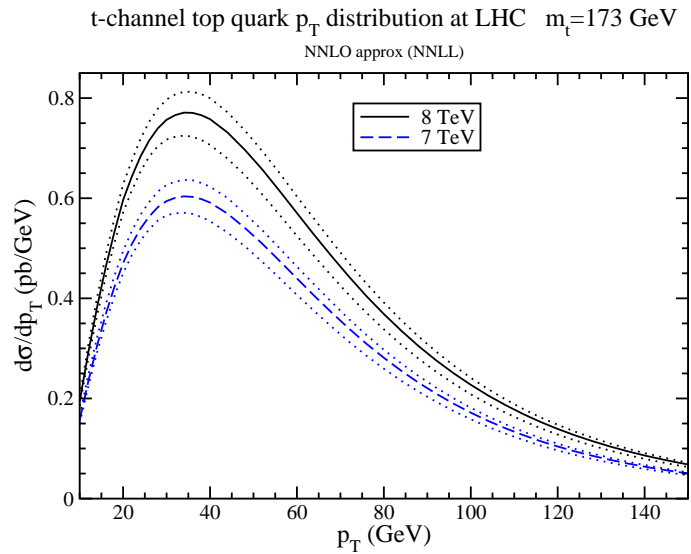}
\caption{The approximate NNLO $t$-channel single-top $p_T$ distributions at 7 and 8 TeV LHC energies. The scale variations are indicated by the dotted lines. }
\label{pttchtoplhcplot}
\end{center}
\end{figure}

We begin with the $t$ channel, which is numerically the dominant single-top process. In Fig. \ref{pttchtoplhcplot} we show the $t$-channel approximate NNLO top-quark $p_T$ distribution \cite{NKtchpt} at 7 and 8 TeV LHC energies. The central results are with $\mu=m_t$ and the variations with renormalization and factorization scales over the interval $m_t/2 < \mu <2m_t$ are also shown. 

\begin{figure}
\begin{center}
\includegraphics[width=9cm]{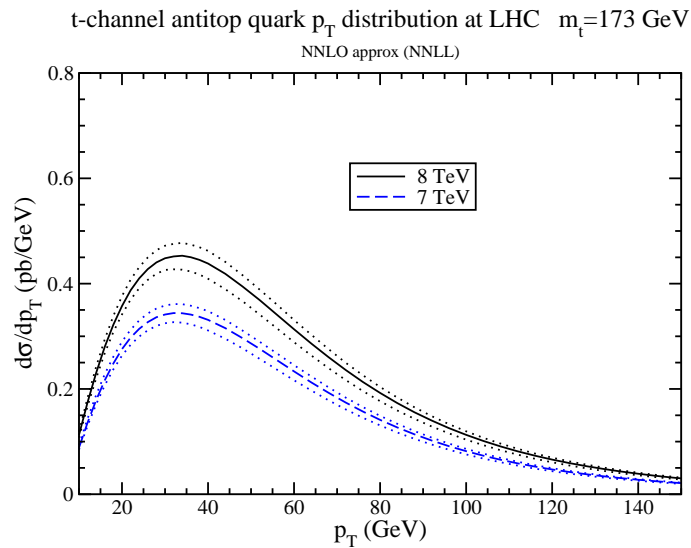}
\caption{The approximate NNLO $t$-channel single-antitop $p_T$ distributions at 7 and 8 TeV LHC energies. The scale variations are indicated by the dotted lines.}
\label{pttchantitoplhcplot}
\end{center}
\end{figure}

The corresponding results for the $t$-channel single-antitop $p_T$ distribution are shown in Fig. \ref{pttchantitoplhcplot}.  

\begin{figure}
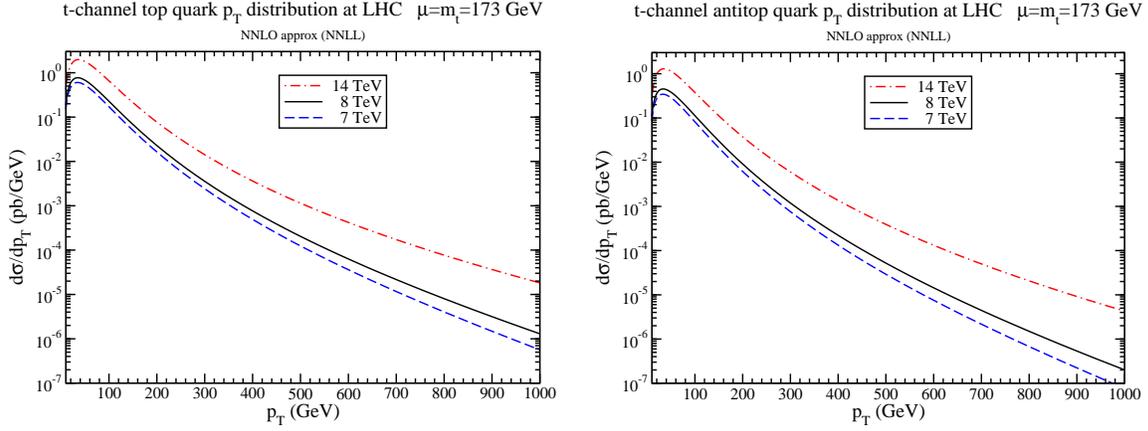

\begin{center}
\includegraphics[width=7.3cm]{pttchtoplhclogplot.eps} 
\hspace{3mm}
\includegraphics[width=7.3cm]{pttchantitoplhclogplot.eps}
\caption{The approximate NNLO $t$-channel single-top (left plot) and single-antitop (right plot) $p_T$ distributions at 7, 8, and 14 LHC energies.}
\label{pttchlhclogplot}
\end{center}
\end{figure}

The plots in Fig. \ref{pttchlhclogplot} show the $p_T$ distributions for $t$-channel single-top (left plot) and $t$-channel single-antitop (right plot) production at 7, 8, and 14 TeV LHC energies over a wider $p_T$ range up to 1000 GeV.

\begin{figure}
\begin{center}
\includegraphics[width=8.5cm]{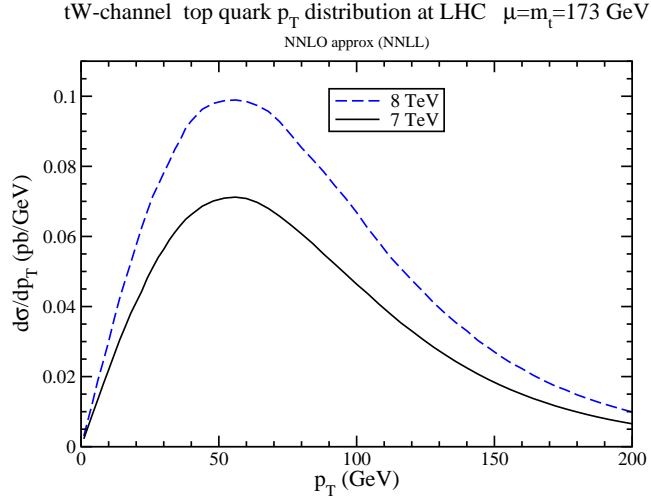}
\caption{The approximate NNLO $tW$-channel single-top $p_T$ distributions at 7 and 8 TeV LHC energies.}
\label{pttWlhcplot}
\end{center}
\end{figure}

Finally, we study the associated production of a top quark with a $W$ boson. This is numerically the second largest single-top process at the LHC. In Fig. \ref{pttWlhcplot} we show the top quark $p_T$ distribution in $tW^-$ production at the LHC at 7 and 8 TeV energies.

\end{document}